\documentclass[12pt]{article}
\usepackage{multicol}
\usepackage{graphicx}

\usepackage{epsfig}

\begin{document}



\title{PS-meson form factors in relativistic quantum mechanics 
and constraints from covariant space-time translations}

\author{%
      Bertrand Desplanques$^{1}$\thanks{{\it E-mail address:}  desplanq@lpsc.in2p3.fr},%
\quad Yubing Dong$^{2,3}$\thanks{{\it E-mail address:} dongyb@ihep.ac.cn}\\%
$^{1}$~Laboratoire de Physique Subatomique et Cosmologie, \\
Universit\'e Joseph Fourier Grenoble 1, CNRS/IN2P3, INPG, France\\
$^{2}$~Institute of High Energy Physics, Chinese Academy of Sciences,\\
Beijing 100049, P. R. China\\
$^{3}$~Theoretical Physics Center for Science Facilities (TPCSF), CAS,\\
Beijing 100049, P. R. China\\
}
\maketitle


\begin{abstract}
The role of  Poincar\'e covariant space-time translations 
is investigated in the case of the pseudoscalar-meson charge form factors 
calculated within a relativistic quantum mechanics framework. 
It is shown that this role extends beyond the standard energy-momentum
conservation, which is accounted for in all works 
based on this general approach. 
It implies constraints that have been largely ignored until now 
but should be nevertheless fulfilled to ensure the full Poincar\'e covariance. 
The violation of these constraints, which is more or less important 
depending on the form of relativistic quantum mechanics that is employed, 
points to the validity of using a single-particle current,
which is generally assumed in calculations of form factors. 
In short, these constraints concern the
relation of the momentum transferred to the constituents to the one 
transferred to the system, which most often differ in relativistic quantum
mechanics while they are equal in field theory. How to account for 
the  constraints, as well as restoring the equivalence 
of different relativistic quantum mechanics approaches in estimating 
form factors, is discussed. It is mentioned that the result so obtained 
can be identified to a dispersion-relation one.
A short conclusion relative to the underlying dynamics is given 
in the pion case.
\end{abstract}

Key words: form factors, pion, relativisty, covariance, translations  \\

PACS: 12.39.Ki, 13.40.Gp, 14.40.Aq

\section{Introduction}
Calculations of form factors in relativistic quantum mechanics (RQM) 
generally imply the choice of a particular hypersurface to describe 
the process under consideration. Among these hypersurfaces, 
those that exhibit some symmetry properties have retained the attention 
and have given rise to various forms of relativity, first considered 
by Dirac \cite{Dirac:1949cp}. The relativistic character of the approach 
supposes that the Poincar\'e algebra, implying rotation, boost 
and space-time translation operators, be satisfied. 
The construction of this algebra within the RQM framework was first performed
by Bakamjian and Thomas for the instant form \cite{Bakamjian:1953kh} 
and extended later on to other forms \cite{Keister:sb}. 
It relies on a mass operator that has to fulfill general conditions but does
not need to be spe\-ci\-fied otherwise. This mass operator can be used in any
form and can involve further relativistic effects that are not required for
ensuring Poincar\'e covariance properties relevant for describing some state. 
Calculations of form factors involve states with different momenta. 
Poincar\'e covariance for such a process implies that form factors
should not depend on the hypersurface chosen to describe the dynamics \cite{Sokolov:1978}, 
which is a matter of convenience. This equivalence of different 
approaches then supposes, generally, to account for the contribution 
of many-particle currents at all orders in the interaction \cite{Sokolov:1978}.
In practice however, calculations are based on a single-particle current 
and, as a result, form factors may depend on the approach. 
Restoring the equivalence of diffe\-rent approaches is a
necessary task prior to any comparison of estimates to measurements.
In the absence of many-particle currents, one could tentatively discriminate 
between the approaches by checking whether they fulfill Poincar\'e 
covariance properties. Most often, the transformations of form factors
under rotations or boosts are discussed. The role of space-time translations 
is limited to the total energy-momentum conservation, 
which is assumed in all cases. In this contribution, we show that 
this conservation property does not exhaust all covariance properties 
of currents under space-time translations (sect. 2). The further constraints 
that they imply \cite{Lev:1993} are considered in detail (sect. 3). 
We describe a way to account for them and discuss their role in restoring the equivalence 
of different RQM approaches, what is made here for pseudoscalar mesons (sect. 4).

Due to a lack of space, we mainly concentrate in the present contribution 
on the essential points underlying our approach for restoring the equivalence 
of different implementations of relativistic quantum mechanics 
in calculating form factors. We refer to refs.
\cite{Desplanques:2010,Desplanques:2009,Desplanques:2008fg} 
for technical details. 

\section{Constraints from transformations of currents under space-time translations}  
Covariant transformations of currents under space-time
translations imply the relation: 
\begin{eqnarray}
J^{\nu}(x) \;({\rm or}\;S(x))=e^{iP \cdot x}\;(J^{\nu}(0) \;({\rm
or}\;S(0)))\;e^{-iP \cdot x}\,.
\label{eq:translat1}
\end{eqnarray}
When it is taken between eigenstates of the total momentum operator, $P^{\mu}$, 
it allows one to factorize the dependence on the space-time coordinate, $x$, 
at the r.h.s.: 
\begin{eqnarray}
<i\;| J^{\nu}(x) \;({\rm or}\;S(x))|\;f>=
e^{i\,(P_i-P_f) \cdot x}\;<i\;|J^{\nu}(0) \;({\rm or} \;S(0))|\;f>.
\label{eq:translat2}
\end{eqnarray}
Combined with the $e^{iq\cdot x}$ function representing an external field 
carrying the momentum $q^{\mu}$, one gets 
the current momentum-energy conservation relation: 
\begin{eqnarray}
(P_f-P_i)^{\mu} = q^{\mu}\, ,
\label{eq:translat3}
\end{eqnarray}
either under the assumption of space-time translation invariance 
or by performing an integration over $x$. At this point, 
we observe that the above relation does not imply any close relation for
$q^{\mu}$ and the momentum transferred to the constituents, $(p_f-p_i)^{\mu}$, 
in contrast to field theory where an equality is expected. 
We also observe that relation (\ref{eq:translat2}) tells nothing 
on whether the current at $x=0$ can be restricted to a single-particle one 
though this is assumed in most works independently of the RQM approach. 

\begin{figure}[htb]
\center{ \epsfig{ file=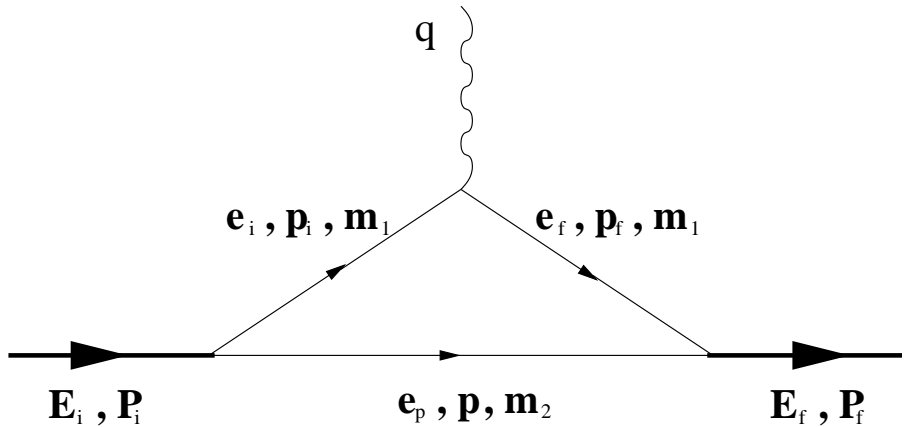, width=12cm}}
\caption{Representation of the interaction with an external probe and
kinematical definitions
\label{fig:probe}}
\end{figure}

The question arises of whether relations stemming from the most general
transformations of currents under space-time translations, 
eq. (\ref{eq:translat1}), implying in particular the vicinity 
of the point $x=0$, can shed some light on the above observations. 
Among the numerous possible relations \cite{Lev:1993}, 
the following ones are especially relevant: 
\begin{eqnarray}
\hspace*{1cm}\Big[P_{\mu}\;,\Big[ P^{\mu}\;,\; J^{\nu}(x)\Big]\Big]=
-\partial_{\mu}\,\partial^{\mu}\,J^{\nu}(x),
\;\;\; 
\Big[P_{\mu}\;,\Big[ P^{\mu}\;,\; S(x)\Big]\Big]=
-\partial_{\mu}\,\partial^{\mu}\,S(x) \, . 
\label{eq:translat4}
\end{eqnarray}
Considering the matrix element of these relations  between eigenstates 
of $P^{\mu}$ and assuming a single-particle current, one should satisfy 
the relation: 
\begin{eqnarray}
<\;|q^2\; J^{\nu}(0) \;({\rm or}\;S(0))|\;>=
<\;|(p_i-p_f)^2\,J^{\nu}(0)\;({\rm or}\;S(0))|\;> \,. 
\label{eq:translat5}
\end{eqnarray}
It is easily seen that this equation cannot be generally fulfilled 
in RQM approaches as, most often, $q^2 \neq (p_i-p_f)^2$ 
(see fig. \ref{fig:probe} for a graphical representation). 
This implies that the assumption of a single-particle current is inconsistent 
with the covariance properties from space-time translations. 
The current, $J^{\nu}(0)\; ({\rm or}\; S(0))$, besides a single-particle 
component, should therefore also contain many-particle components 
which, until now, have been ignored. There is however one exception. 
In the standard front-form case ($q^+=q \cdot \omega =0$), it turns out 
that the relation $q^2=(p_i-p_f)^2$ is fulfilled, implying 
that eq. (\ref{eq:translat5}) is always satisfied, while being limited to a
single-particle current.

%
\section{Implementation of the constraints}
It is generally expected that many-particle currents at all orders 
in the interaction are required to fulfill the full Poincar\'e covariance
properties, possibly allowing one to restore the equivalence of different
approaches for calculating form factors. As, for numerical reasons, 
the simplest two-particle currents  are rarely considered
\cite{Desplanques:2009b,Desplanques:2003nk}, 
the above task seems to be out of reach {\it a fortiori}. 
However, the many-particle currents of interest here represent a specific
subset, with a well defined role, namely accounting for constraints from
covariant space-time translations. Moreover, with this respect, 
the front-form approach is consistent with a single-particle current. 
One can therefore infer that there may be some trick allowing one 
to sum up the contributions of the many-particle currents 
so that the single-particle current structure 
appropriate for the front-form approach be recovered.
The trick we found is suggested by examining expressions of form factors. 
They show that the factors multiplying $Q$ 
in different approaches, 1 and $\frac{2e_k}{M}$ in the most
striking cases, differ by terms that have typically an
interaction character \cite{Desplanques:2003nk} 
and are a signature of describing physics on different underlying hypersurfaces. 
The idea is to multiply Q by a factor $\alpha$ so that to account 
for the further interaction currents expected to restore 
the equivalence of different approaches. The factor $\alpha$ 
is determined  by requiring that the squared momentum transferred 
to the system, $q^2$, be equal to the one for the constituents,  
denoted $``(p_i-p_f)"^2$. The equation to be solved is typically given by:
\begin{eqnarray}
q^2&=&
``[(P_i\!-\!P_f)^2+2\, (\Delta_i \!-\! \Delta_f)\;  (P_i\!-\!P_f) \cdot \xi
+(\Delta_i \!-\!\Delta_f)^2 \;\xi^2]"
\nonumber \\
&=&\alpha^2q^2-2\,\alpha\; ``(\Delta_i \!-\!\Delta_f)" \;q \cdot \xi
 +``(\Delta_i \!-\!\Delta_f)^2"\;\xi^2 \,,
 \label{eq:constraint1}
\end{eqnarray}
where $\Delta$, which represents an interaction effect, also depends 
on $\alpha$. Explicit expressions of $\alpha$ can be found for different forms.  
Expressions for form factors, taking into account the effect of constraints
motivated by space-time translation properties, 
can then be obtained  \cite{Desplanques:2010,Desplanques:2009}. 
The many-particle character of corrections to the current at all orders 
of the interaction could be checked by expanding these expressions 
in terms of  $\Delta$. It is worthwhile  to  notice that, 
for the standard front-form where $\xi^2=0$, $q\cdot\xi\;({\rm or}\; q^+)=0$, 
the factor $\alpha$ is equal to 1 and, therefore, results for the form factors 
are unchanged. Moreover, the expression of the single particle-current 
in other forms, before accounting for the constraints, may slightly differ from 
what is  expected in an impulse approximation 
but the difference involves effects that are typically interaction ones. 
Its choice ensures that results for form factors are Lorentz invariant.
\section{Restoration of the equivalence: numerical illustration and comments}
\begin{figure}[htb]
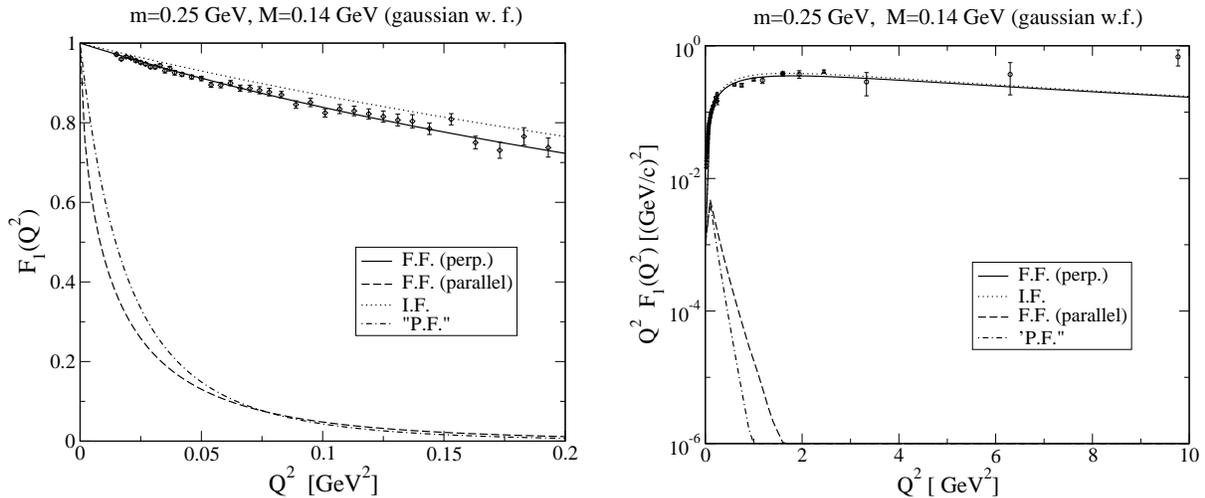

\mbox{ \epsfig{ file=pions.eps, width=7.6cm}}
\hspace*{0.4cm}
\mbox{\epsfig{ file=pionss.eps, width=7.6cm}}
\caption{Pion charge form factor at low and intermediate $Q^2$, without and with the
effect of constraints considered in this work. In the last case, all curves
coincide with the standard front-form one (F.F. (perp.)).   
\label{fig:charge}}
\end{figure} 
To illustrate effects of the restoration of properties related to space-time 
translations, we calculated the charge form factor for both the pion and kaon
mesons using a Gaussian wave function which approximately accounts for the
confinement interaction (string tension equal to 0.2 GeV$^2$). 
The quark mass is fixed by fitting the meson decay constant.
We present in fig. \ref{fig:charge} results for the pion, 
at low $Q^2$ in the left panel and at intermediate $Q^2$ in the right panel.
The left panel shows the sensitivity to the charge radius. 
In the right panel, the form factor has been multiplied by $Q^2$ 
as often done on the basis that the corresponding product could tend 
to a constant asymptotically (up to log terms). We nevertheless stress 
that the two-particle contribution to the current which could reproduce 
this behavior \cite{Desplanques:2010,Desplanques:2009b} 
is not included in the results presented here.
The various curves shown in the figure correspond to different forms 
or different momentum configurations and are obtained for the Breit frame 
(F.F. (perp.): standard front form with $q^+=0$; 
F.F. (parallel): front form with $\vec{q}$ parallel to the front orientation;
I.F.: standard instant form \cite{Bakamjian:1953kh};
``P.F.": ``instant form with the symmetry properties of the Dirac point form" 
\cite{Bakamjian:1961}). We do not include results for the Dirac point form
(hyperboloid hypersurface), which fall between the two front-form curves
\cite{Desplanques:2004rd}.

Examination of results without the effect of constraints considered in this
work evidences striking features. While the standard front-form 
and instant-form results are close to each other, 
they show tremendous discrepancies with the other two curves. 
The fast fall off of form factors at low  $Q^2$ in the last cases 
reflects a somewhat paradoxical $1/M^2$ dependence 
of the pion squared charge radius. Though there was no intent 
to make a detailed  comparison to measurements, it is clear 
that the two first approaches do relatively well, despite the corresponding
results are not Lorentz invariant {\it a priori} while the ``P.F." one, 
which evidences this property, does very badly. 
After incorporating the effect of constraints, all curves are found to coincide
with the standard front-form one, which is not changed.
The paradox of a charge radius tending to infinity while the mass 
of the system goes to zero (or the interaction is increased) has disappeared, 
suggesting that the effect of constraints we considered corrects for the
breaking of some symmetry related to space-time translations. 
For the sake of the illustration, we presented results obtained 
in the Breit frame but we could choose any frame as well. 
We stress that the results accounting for the constraints are independent 
of both the frame and the orientation of the hyperplane $\xi^{\mu}$. 
These properties can be readily checked by looking at the expression 
that the pion charge form factor takes after making a change of variables 
and integrating over one of them:
\begin{eqnarray}
\hspace*{-0.8cm}F_1(Q^2) =\frac{1}{N} \!\int\! d\bar{s} \;  d(\frac{s_i\!-\!s_f}{Q}) \; 
 \phi(s_i) \; \phi(s_f)
\;\frac{ 2\sqrt{s_i\,s_f}
\;\theta(\frac{s_i\,s_f}{D} -m^2)}{D\sqrt{D}} \, .
\label{eq:disp}
\end{eqnarray}
This expression agrees with a dispersion-relation one found prior 
to this work by Melikhov \cite{Melikhov:2001zv} but disagrees 
by a factor $(s_i\!+\!s_f\!+\!Q^2)/(2\sqrt{s_is_f})$ with the one 
obtained by Krutov and Troitsky \cite{Krutov:2001gu}. The discrepancy factor 
in this case is the same as for scalar constituents \cite{Desplanques:2008fg}.

\begin{figure}[htb]
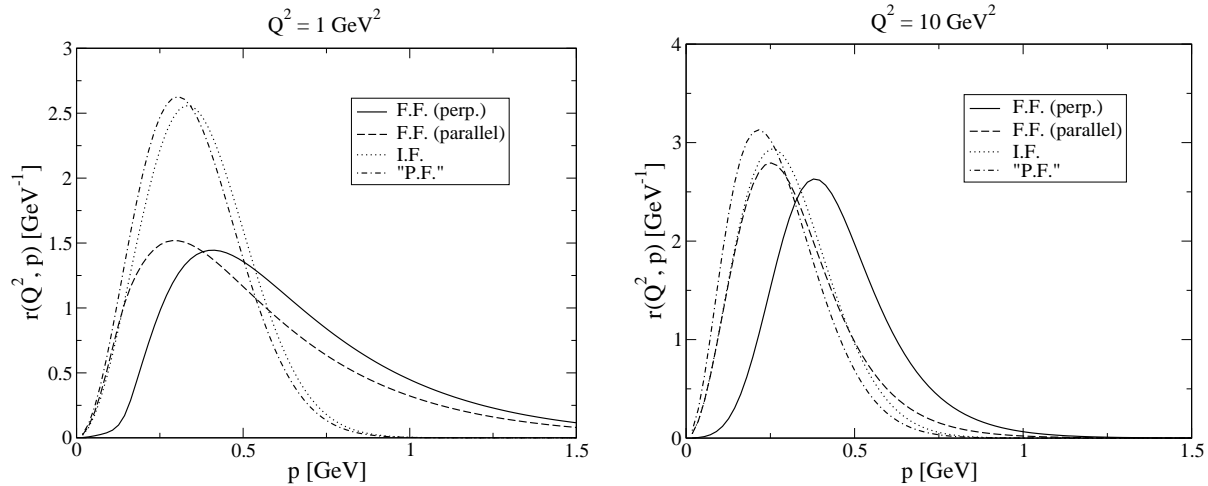

\mbox{ \epsfig{ file=ffx1.eps, width=7.7cm}}
\hspace*{0.4cm}
\mbox{\epsfig{ file=ffx10.eps, width=7.6cm}}
\caption{Representation of the integrand in different approaches as a function
of the spectator momentum, $p=|\vec{p}|$ (see ref. \cite{Desplanques:2010} for a precise definition of the quantities 
shown in the figure).   
\label{fig:dependonp}}
\end{figure} 
We would like to mention that getting for all approaches the same expression 
of the form factor in terms of the Mandelstam variables $s_i, \;s_f$, 
eq. (\ref{eq:disp}), does not imply that the expression in terms of the
spectator momentum, $\vec{p}$, is the same. This is illustrated 
in fig. \ref{fig:dependonp} where we show the integrands in terms of the
quantity $p=|\vec{p}|$ corresponding to results presented 
in fig. \ref{fig:charge} at two momentum transfers $Q^2= 1$ and $Q^2= 10$
GeV$^2$. The change of variables that allows one to make the integration over
one of them and reduce a 3-dimensional integration to a 2-dimensional one 
is specific of each approach and is non-trivial. It allows one to concentrate
all the explicit dependence of the integrand on the frame or on the front 
orientation in a new variable that can be integrated over, 
leaving only  $s_i$ and $s_f$ as integration variables.

The pseudoscalar mesons, besides a charge form factor,  have a Lorentz scalar
form factor. Results in this case are qualitatively very similar to the charge form
factor ones. Charge form factors were calculated taking into account that the
constituents have the spin structure appropriate to quarks (spin-1/2). They of
course differ from those for scalar constituents \cite{Desplanques:2008fg} but, roughly, 
we observe the same qualitative features (the largest effects involve the wave
function). Charge form factors have also been calculated 
for the kaon meson \cite{Desplanques:2010}. The effect of constraints is not as large as for the pion
case. The reason is to be looked for in the kaon mass. It is reminded that the
effect, in the most striking cases, involves terms $2e_k/M$, which is
smaller for the kaon than for the pion.

\section{Conclusion}
We considered in this work properties related to Poincar\'e covariant
space-time translations in RQM approaches for the calculation of form factors 
of pseudoscalar mesons. Apart from the standard front form with $q^+=0$, 
all other RQM approaches need to be completed for the many-particle currents 
that the above properties, generally ignored, imply. When this is done, it is found  
that discrepancies between different RQM approaches 
for calculating form factors of pseudoscalar mesons can be removed, 
showing that the role of space-time translations extends 
beyond the standard energy-momentum conservation. 
It is also found that these results could coincide with those 
of a dispersion-relation approach. Results are qualitatively 
very similar to those obtained in a previous work with scalar constituents 
\cite{Desplanques:2008fg}.  
Altogether, all aspects of the Poincar\'e 
group (rotations, boosts but also space-time translations) are essential 
in getting reliable estimates of form factors as far as the implementation of
relativity is concerned. While this important result is obtained by considering
many-particle currents, we would like to emphasize that these ones represent a
mi\-ni\-mal subset that allows one to fulfill expected transformation properties of
observables when the generators of the Poincar\'e group are applied to them. It
is also noticed that the invariance of form factors under Lorentz
transformations alone is by far not sufficient to guarantee the validity of an
approach, as evidenced by the ``P.F."  example. This observation raises another
question. The fast fall off of the pion form factor in this last approach is
obtained in a truncated front-form field-theory calculation
\cite{Simula:2002vm,Bakker:2001pk,deMelo:2002yq}, where the effect is
interpretated as due to missing the contribution of zero modes related to the
breaking of the rotational symmetry, as well as in an instant form with an
infinite-momentum configuration, where, instead, the rotational symmetry is
fulfilled. Results presented in this work would rather suggest that the fast
fall off of the pion charge form factor in some calculations is due to
the fact that, besides breaking rotational or boost invariance, they
simultaneously miss part of transformation properties under space-time translations. 
This aspect should deserve specific studies.
We mentioned that restoring the equivalence of different approaches was
requiring the contribution of a subset of many-particle currents. 
As an example of  further currents, we could mention those allowing one 
to reproduce the expected asymptotic behavior of the pion charge form factor, 
which was recently determined in a RQM framework \cite{Desplanques:2009b}.

Having determined a set of results that are independent of the chosen
implementation of re\-la\-tivity, one can proceed to their comparison with
measurements. This comparison can provide information on the mass operator 
that can be taken to be the same in all approaches. 
This operator possibly involves relativistic effects that differ from those 
ensuring the equivalence of approaches for the calculation of form factors.
Taking the standard front-form results as representative 
of these common results, we find that the calculated charge form factor 
does rather well in the pion case. Results for the kaon charge form factor 
are not so good but, due to various uncertainties, it is difficult 
to claim that there is a real discrepancy. 
In any case however, there are many reasons to consider the above results 
for the charge form factors as suspicious, especially in the pion case 
where there is a relatively good agreement. 
The calculation misses the whole physics related to the one-gluon exchange 
in both the interaction entering the mass operator and in the currents. 
Their contributions were considered in various works  but, 
whatever the approach, accounting for them tends to overshoot measurements 
in the intermediate $Q^2$ range. 
Part of the solution to this problem probably requires to take into account
in the determination of the mass operator corrections that correspond 
to retardation effects in a field-theory approach. 
These effects decrease the strength of the one-gluon exchange
interaction obtained in the instantaneous approximation. Similar effects 
were found to be relevant for a system made of scalar constituents 
\cite{Desplanques:2008fg}.
Another part of the solution to the overshooting problem may require 
the introduction  of some quark form factors 
\cite{Cardarelli:1994,Cardarelli:1995dc,deMelo:2007,Gross:2008}. 
Its choice should however take into account that part of the effect underlying
the vector meson dominance phenomenology is already included. 
Moreover, its effect should be consistent with the expectation 
of a vanishing in the limit of a zero value for the QCD coupling, $\alpha_s$. 
With this respect, the approach developed in ref. \cite{deMelo:2007} 
could be more appropriate.
 
The present work was extending to spin-1/2 constituents a previous one 
for scalar constituents. Further work along present lines would involve 
inelastic and time-like processes as well as  non-zero spin systems
such as meson resonances or the nucleon and its resonances. 
This task could require more elaboration. Let's mention that 
in the nucleon case, large effects from the constraints considered 
in this work are expected too though not as large as for the pion.  
The factor that determines the size of these effects is roughly 
a half of what it is in the pion case, where the corresponding factor, 
$2e_k/M$, can be as large as 6 at low  $Q^2$. 
Smaller effects are expected for resonances however.

{\bf Acknowledgments}\\
This work is partly supported by the National Sciences Foundations of China
under grants No. 10775148,10975146 (Y.B.). The author is also grateful to the
CAS for grant No KJCX3-SYW-N2. B.D. would like to thank IHEP and  LPSC 
for offering hospitality allowing him to achieve this work. He is also very
grateful for support from valencian colleagues, which allowed him to attend 
the LC2010 conference.

\end{document}